\documentclass[reprint,prb,superscriptaddress,aps
]{revtex4-1}

\usepackage{bm}%
\usepackage[colorlinks=true,linkcolor=blue]{hyperref}%
\expandafter\ifx\csname package@font\endcsname\relax\else
 \expandafter\expandafter
 \expandafter\usepackage
 \expandafter\expandafter
 \expandafter{\csname package@font\endcsname}%
\fi
\usepackage{graphicx}
\usepackage{dcolumn}

\begin{document}

\preprint{APS/123-QED}

\title{Wave transmission and its universal fluctuations in one-dimensional systems with L\'evy-like disorder: Schr\"odinger, Klein-Gordon and Dirac equations}

\author{Anderson L. R. Barbosa}
\email{anderson.barbosa@ufrpe.br}
\affiliation{Departamento de F\'{\i}sica, Universidade Federal Rural de Pernambuco, 52171-900, Recife, PE, Brazil}

\author{Jonas R. F. Lima}
\email{jonas.lima@ufrpe.br}
\affiliation{Departamento de F\'{\i}sica, Universidade Federal Rural de Pernambuco, 52171-900, Recife, PE, Brazil}

\author{Luiz Felipe C. Pereira}
\email{pereira@df.ufpe.br}
\affiliation{Departamento de F\'{\i}sica, Universidade Federal de Pernambuco, 50670-901, Recife, PE, Brazil}

\date{\today}

\begin{abstract}
We investigate the propagation of waves in one-dimensional systems with L\'evy-type disorder.
We perform a complete analysis of non-relativistic and relativistic wave transmission submitted to potential barriers whose width, separation or both follow L\'evy distributions characterized by an exponent $0 < \alpha <1$. 
For the first two cases, where one of the parameters is fixed, non-relativistic and relativistic waves present anomalous localization, $\langle T \rangle \propto L^{-\alpha}$.
However, for the latter case, in which both parameters follow a L\'evy distribution, non-relativistic and relativistic waves present a transition between anomalous and standard localization as the incidence energy increases relative to the barrier height.
Moreover, we obtain the localization diagram delimiting anomalous and standard localization regimes, in terms of incidence angle and energy.
Finally, we verify that transmission fluctuations, characterized by its standard deviation, are universal, independent of barrier architecture, wave equation type, incidence energy and angle, further extending earlier studies on electronic localization.

\end{abstract}


\pacs{Valid PACS appear here}
\maketitle


\section{Introduction}

Wave localization has been an import subject of research both theoretically and experimentally \cite{Kramer_1993,RECAMI2009235,JOUR,https://doi.org/10.48550/arxiv.2106.13892}. 
The goal is to understand the localization behavior of classical and quantum waves when crossing a disordered region. 
The first step was given by P. W. Anderson in his seminal work \cite{Anderson1958} which introduced the concept of strong localization in the quantum electronic wave context, and was later extended to mechanical and electromagnetic waves \cite{JOUR, ultrasound}. 
Over the last few years, the localization of quantum waves submitted to random disorder distribution has been investigated using the Schr\"odinger, the Klein-Gordon and the Dirac equations \cite{PhysRevLett.113.233901,PhysRevB.85.104201, BARBOSA2020114210,PhysRevB.100.104201}.

However, several phenomena in nature and human activity are actually described by long-tail distributions, known as L\'evy distributions \cite{Song1018,PhysRevLett.78.3864,PhysRevLett.71.3975,PhysRevE.95.032315,Nature453,PhysRevE.91.032112,PhysRevB.98.235144,PhysRevA.85.035803,vanLoevezijn:96,2040-8986-17-10-105601,doi:10.1080/09500340.2017.1310317}.
Thus, understanding the consequences of L\'evy-like disorder in condensed matter systems became an active research topic \cite{PhysRevLett.117.126801,0295-5075-92-5-57014,PhysRevB.24.5698,0305-4470-36-12-322,PhysRevB.85.235450,PhysRevB.88.205414,PhysRevLett.117.046603,0957-4484-28-13-134001,PhysRevB.97.075417,PhysRevE.96.062141,PhysRevE.93.012135,Lima2019,Razo-Lopez2020,PhysRevE.105.024131}.
A L\'evy distribution is characterized by a probability density $\rho(w)$ of a random variable $w$, which has a power-law tail \cite{Nature453,PhysRevE.96.062141,PhysRevB.98.235144}. 
In general, the respective probability density can be written as 
\begin{equation}
\rho(w) \propto \frac{1}{w^{1+\alpha}},\label{Levy}
\end{equation}
where $0 < \alpha < 2$. 
If $0 < \alpha < 1$, the first and second moments of $\rho(w)$ diverge because of the heavy tails, while for $1 \le \alpha < 2$ only the second moment diverges. 
For instance, Ref. [\onlinecite{Nature453}] developed a glass disorder sample formed by glass microspheres, whose diameter follow a L\'evy-type distribution, which was used to show that the light transmission decays as a power-law. 
Ref. [\onlinecite{Kohno2004}] observed a non-periodic diameter modulation in SiC nanowires, and verified that the diameter fluctuations followed a power-law distribution.
Motivated by Refs. [\onlinecite{Kohno2004}], the authors of Ref. [\onlinecite{0295-5075-92-5-57014}] performed an analytical investigation of quantum electronic waves submitted to potential barriers where the spacing between barriers followed a L\'evy distribution.

Quantum electronic waves submitted to typical one-dimensional random disorder show  standard localization, which means that the average transmission decays exponentially with the system length $L$, $\left\langle T \right\rangle \propto \exp{\left(-L/\lambda\right)}$ and consequently $\left\langle - \ln T \right\rangle \propto L/\lambda$, $\lambda$ being the localization length.
Meanwhile, Refs. [\onlinecite{0295-5075-92-5-57014,PhysRevA.85.035803,PhysRevLett.113.233901}] showed that for L\'evy-like one-dimensional disorder the quantum electronic waves follow anomalous localization with a power-law decay
\begin{equation}
\left\langle T \right\rangle \propto L^{-\alpha},
\label{TL}
\end{equation}
while the average minus transmission logarithm behaves as  
\begin{equation}
\left\langle - \ln T \right\rangle \propto L^{\alpha},
\label{lnLa}
\end{equation}
when $0<\alpha<1$.

These results have been followed by other works studying non-relativistic wave transmission submitted to potential barriers with L\'evy distribution \cite{PhysRevLett.117.046603,PhysRevLett.123.195302}. 
Recently, we studied the effect of L\'evy-like one-dimensional disorder on the transmission of relativistic waves \cite{Lima2019}. 
We showed that when the width of barriers and spaces between them follow a L\'evy distribution, the transmission presents a transition from anomalous to standard localization as the incidence energy increases. 
Studies in which the waves are described by the Klein-Gordon equation in the presence of a L\'evy-like disorder have not yet been presented in the literature.

\begin{figure}
\centering
\includegraphics[width=1.0\linewidth]{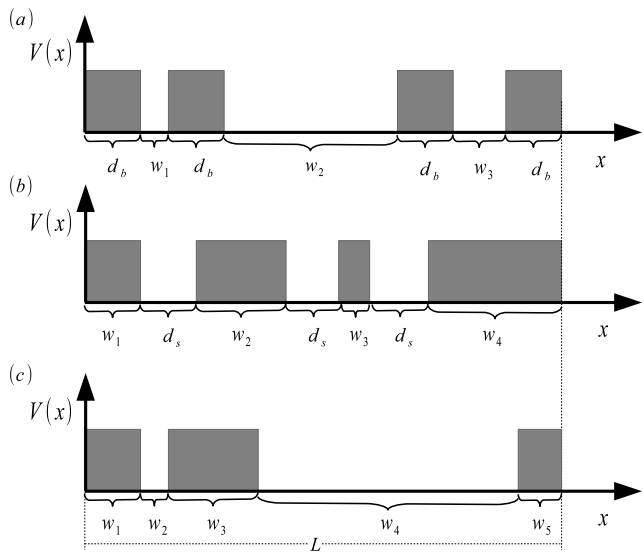}
\caption{Schematic representation of three distinct potential barrier architectures for a fixed system length $L$. (a) The width of the barriers $d_b$ is fixed while the separation  between them $w_i$ follows a L\'evy distribution. (b) The width of the barriers $w_i$ follows a L\'evy distribution while their separation $d_s$ is fixed. (c) The width of  barriers and their separation $w_i$ follow a L\'evy distribution. The total number of barriers and spaces is defined by $L$, all barriers have the same height.}
\label{potencial}
\end{figure}

In this context, here we employ the transfer matrix method to develop a complete analysis of non-relativistic and relativistic wave transmission submitted to potential barriers following L\'evy distributions. 
We address three distinct barrier architectures, which are schematically shown in Fig. \ref{potencial}: (a) the width of the barriers $d_b$ fixed while the separation between them $w_i$ follows a L\'evy distribution; (b) the width of the barriers $w_i$ follows a L\'evy distribution while their  separation $d_s$ is fixed; and (c) the width $w_i$ of the barriers and the separation between them follow a L\'evy distribution. 

For the first two architectures, non-relativistic and relativistic waves present anomalous localization for all incidence energy and angle when $0<\alpha<1$. 
However, for the latter architecture both non-relativistic and relativistic waves present a transition between anomalous and standard localization as the incidence energy increases for $0<\alpha<1$. 
Moreover, we obtain the localization diagram delimiting anomalous and standard localization regimes, in the form of an incidence angle versus incidence energy diagram. 
Finally, we verify that transmission fluctuations (characterized by its standard deviation) are universal, independent of barrier architecture, wave equation type, incidence energy and angle, further extending earlier studies on localization of electrons and phonons \cite{Lee1985,Giordano1988,Nishigushi1993}.

\section{Schr\"odinger equation}
\label{sec:schrodinger}

\subsection{Model}

The Schr\"odinger equation for non-relativistic quantum particles in two-dimensions subject to  one-dimensional potential barriers can be written as
\begin{equation}
    (\partial^2_x + \partial_y^2) \psi(x,y) = -\frac{2m}{\hbar^2}[E-V(x)]\psi(x,y).
   \label{seq}
\end{equation}
Due to momentum conservation in the $y$ direction, it is possible to write $\psi(x,y)=\psi(x)e^{ik_yy}$, such that Eq. (\ref{seq}) becomes
\begin{equation}
      \partial^2_x \psi(x) = -\frac{2m}{\hbar^2}\left(E-V(x)-\frac{\hbar^2k_y^2}{2m}\right)\psi(x).
      \label{scheqx}
\end{equation}
Inside the $j$th region of the superlattice $V(x)$ is constant, so the solution for this equation is
\begin{equation}
    \psi(x) = \left\{  \begin{array}{c}
        Ae^{ik_xx} + Be^{-ik_xx}, \quad \text{regions with} \quad V(x) = 0 \\
        A^{\prime}e^{ik_x^{\prime}x} + B^{\prime}e^{-ik_x^{\prime}x},  \quad \text{regions with} \quad  V(x) = V
    \end{array}\right.
\end{equation}
where
\begin{equation}
    k_x = \frac{\sqrt{2mE-\hbar^2k_y^2}}{\hbar}
\end{equation}
and
\begin{equation}
    k_x^{\prime} = \frac{\sqrt{2m(E-V)-\hbar^2k_y^2}}{\hbar}.
\end{equation}
It is important to mention that in order to obtain the transmission as a function of the incidence angle $\theta_0$, one has to write
\begin{equation}
    k_y = \sin \theta_0 \frac{\sqrt{2mE}}{\hbar}.
\end{equation}

Now, considering the continuity of $\psi(x)$ and its derivative at the interfaces between $N$ regions of a superlattice, it is possible to write a transfer matrix connecting the amplitude of the waves in the incidence region with the waves in the exit region, which gives us
\begin{equation}
    \left(\begin{array}{c}
        A   \\
        B  
    \end{array} \right) = M 
     \left(\begin{array}{c}
        C   \\
        0  
    \end{array} \right),
    \label{tmeq}
\end{equation}
where $A, B$ and $C$ are the amplitudes of the incident, reflected and transmitted waves, respectively, such that 
\begin{equation}
    M = I_1P_1I_2P_2I_1P_3I_2P_4...I_1P_NI_2.
\end{equation}
Here, we have that
\begin{equation}
    I_i =\frac{1}{t_i} \left(\begin{array}{cc}
       1  & r_i \\
       r_i  & 1
    \end{array} \right),
\end{equation}
with $i=1,2$ and $r_1=(k_x-k_x^{\prime})/(k_x+k_x^{\prime})$, $r_2=(k_x^{\prime}-k_x)/(k_x+k_x^{\prime})$, $t_1=2k_x/(k_x+k_x^{\prime})$ and $t_2=2k_x^{\prime}/(k_x+k_x^{\prime})$. We have also that
\begin{equation}
    P_j = \left(\begin{array}{cc}
       e^{ik_x^jw_j}  & 0 \\
       0  & e^{-ik_x^jw_j}

    \end{array} \right),
\end{equation}
where $w_j$ is the $j$th region of the superlattice, which can be a potential barrier or an empty space region.
From Eq. (\ref{tmeq}) we can obtain the transmission as the ratio between the norm of the amplitude for incident and transmitted waves
\begin{equation}
    T = \frac{|C|^2}{|A|^2} = \frac{1}{|M_{1,1}|^2}.
\end{equation}

Following the model described above, for each one of the barriers architectures illustrated in Fig. \ref{potencial}, we calculated the average wave transmission $\langle T \rangle$ and 
the average of minus its logarithm  $\langle -\ln T \rangle$, performing  $10^4$ independent realizations.
Furthermore, we characterize transmission fluctuations by its standard deviation defined as 
\begin{equation}
    \sigma = \sqrt{\langle T^2 \rangle-\langle T \rangle^2}.
\label{SD}
\end{equation}

\begin{figure}
\centering
\includegraphics[width=0.9\linewidth]{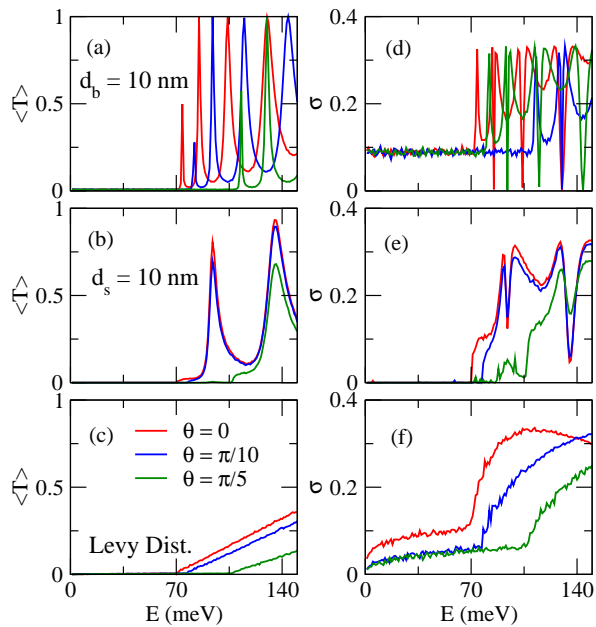}
\caption{The top panels (a,d) are the lattice architecture in Fig. \ref{potencial} (a), middle panels (b,e) correspond to Fig. \ref{potencial}(b) and down panels (c,f) to Fig. \ref{potencial} (c). Transmission average as a function of energy for the Schr\"odinger equation for (a) barrier width $d_b = 10$ nm, (b) barrier separation $d_s = 10$ nm and (c) L\'evy distribution for both. In all calculations we used $\alpha = 0.5$, $L=10$ $\mu$m and potential height $V = 70$ meV  for different incidence angles $\theta_0 = 0, \pi/10$ and $\pi/5$. Panels (d,e,f) are the respective transmission fluctuation.}
\label{Imagem1}
\end{figure}

\subsection{Results}

In Fig. \ref{Imagem1} (a), (b) and (c) we show the average transmission $\left\langle T \right\rangle$, and in panels (d), (e) and (f) the respective standard deviation, as a function of energy for the three barriers architectures illustrated in Fig. \ref{potencial} (a), (b) and (c), respectively.
The data has been averaged over $10^4$ independent realizations with fixed potential height $V=70$ meV, $L=10$ $\mu$m and L\'evy distribution exponent $\alpha=0.5$.

\begin{figure*}
\centering
\includegraphics[width=0.8\linewidth]{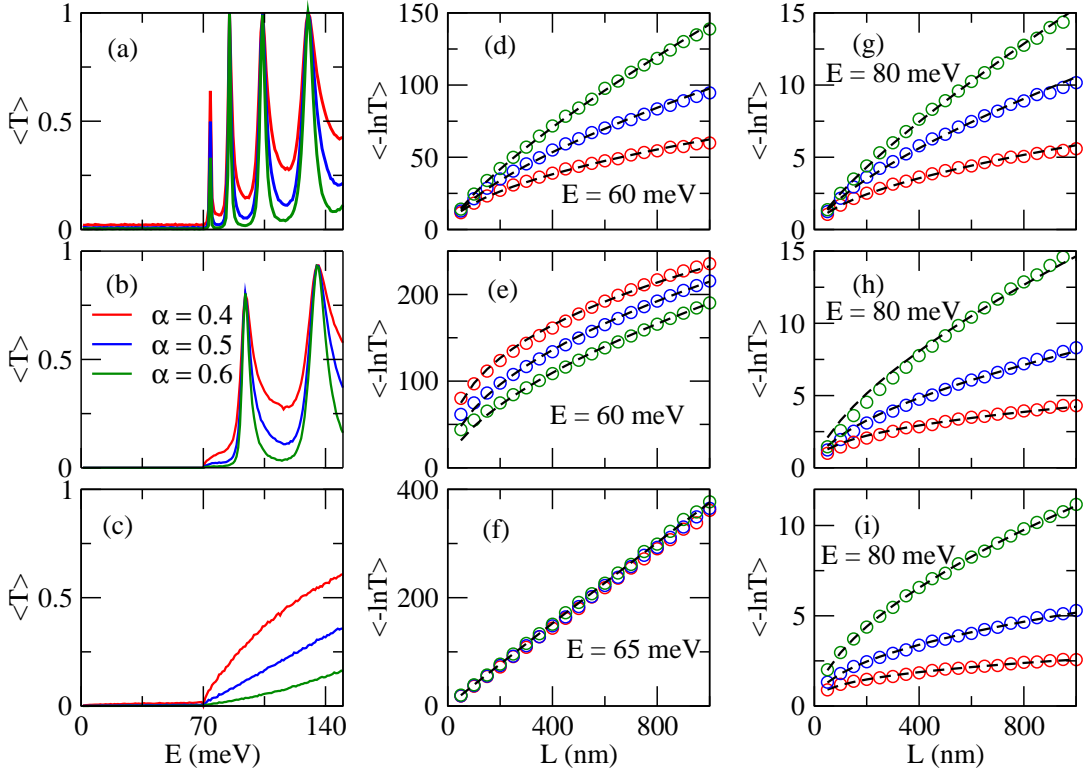}
\caption{Top panels (a,d,g) are for the lattice architecture in Fig. \ref{potencial} (a), middle panels (b,e,h) correspond to Fig. \ref{potencial} (b) and down panels (c,f,i) correspond to Fig. \ref{potencial} (c). Transmission average as a function of energy with $L=10$ $\mu$m, $V = 70$ meV and $\theta_0 = 0$, where (a) barrier width $d_b = 10$ nm, (b) barrier separation $d_s = 10$ nm, and (c) L\'evy distribution for both. $\left\langle -\ln T \right\rangle$ as a function of $L$ for fixed incidence energy when $E < E_c$ (d,e,f) and $E > E_c$ (g,h,i). In panels (d,e,g,h,i) the dashed lines are fitted using Eq. (\ref{lnLa}), while in (f) it is a linear fit.} 
\label{Imagem2}
\end{figure*}

For normal incidence, $\theta_0=0$, the average transmission is close to zero when the energy is lower than the potential barrier height ($E<V$), and it is finite when $E> V$, as shown on the left panels of Fig. \ref{Imagem1}.
Although the average transmission is close to zero when $E<V$, its standard deviation does not vanish for the two architectures in which the separation between barriers is not fixed, Fig. \ref{Imagem1} (d) and (f).
This indicates that when the barrier separation follows a L\'evy distribution, most realizations contribute zero to the transmission, but some contribute with a high value, leading to a nonzero standard deviation. 
On the other hand, when $E> V$ the first two architectures show typical transmission oscillations caused by wave interference, Fig. \ref{Imagem1} (a) and (b), while in the last architecture the average transmission increases monotonically, Fig. \ref{Imagem1} (c).
The standard deviation presents an outstanding feature: its maximum value is the same for the three architectures, which indicates a universal behavior.

In the case of oblique incidence angles $\theta_0=\pi/10$ and $\pi/5$, the behavior of the average transmission and standard deviation are qualitatively equivalent to the $\theta_0=0$ case, as seen in  Fig. \ref{Imagem1}. 
However, the energy at which the average transmission is no longer null undergoes a shift. 
This critical energy $E_c$, can be obtained from the relation
\begin{equation}
    \theta_0=\arcsin \sqrt{1-V/E_c}.\label{SH}
\end{equation}
Notice that for $E_c=V$, we obtain $\theta_0=0$, as expected.

Fig. \ref{Imagem2} (a), (b) and (c) presents the average transmission as a function of energy for $V=70$ meV, $L=10$ $\mu$m and three different L\'evy distribution exponents: $\alpha=0.4, 0.5$ and $0.6$.
In this case we consider normal incidence only, such that $\theta_0=0$ and the critical energy is $E_c = V = 70$ meV in all panels.
Furthermore, they show a similar behavior when compared to the case considered in Fig. \ref{Imagem1}.
In order to understand the localization process for a non-relativistic wave through the potential barrier architectures considered here, we also plot  $\left\langle - \ln T \right\rangle$ as a function of length $L$.
In the center panels (d-f) we have $E<E_c$ while the panels on the right side (g-i) consider $E>E_c$.
The numerical data (circles) in (d) and (e) can be well fitted by Eq. (\ref{lnLa}) (dashed lines), which means an anomalous localization behavior. 
Meanwhile, in panel (f),  $\left\langle - \ln T \right\rangle$ increases linearly with $L$, which indicates standard localization. 
In (g-i) the numerical data is also fitted by Eq. (\ref{lnLa}), which also means anomalous localization.

Therefore, from Fig. \ref{Imagem2}, we can conclude that the barrier architectures shown in Fig. \ref{potencial} (a) and (b) induce anomalous localization for both $E<E_c$ and $E>E_c$, which is in agreement with Ref. [\onlinecite{0295-5075-92-5-57014}]. 
However, for the barrier architecture of Fig. \ref{potencial} (c) we have standard localization for $E<E_c$ and anomalous localization for $E>E_c$.
This standard-to-anomalous localization transition has not been reported previously for non-relativistic waves.

\section{Klein-Gordon equation}
\label{sec:KG}

\subsection{Model}

The Klein-Gordon equation in two dimensions for zero-spin bosons in a superlattice with the potential barriers shown in Fig. \ref{potencial} is given by
\begin{equation}
   (\partial^2_x + \partial_y^2) \psi(x,y) = -\frac{1}{\hbar^2 c^2}([E-V(x)]^2-m^2c^4)\psi(x,y).
   \label{kgeq}
\end{equation}
As in the previous case, we can write $\psi(x,y)=\psi(x)e^{ik_yy}$, and thus Eq. (\ref{kgeq}) becomes
\begin{equation}
    \partial_x^2 \psi(x) = -\frac{1}{\hbar^2 c^2}([E-V(x)]^2-\hbar^2 c^2 k_y^2-m^2c^4)\psi(x).
    \label{kgeqx}
\end{equation}
The equation above has the same form as Eq. (\ref{scheqx}), and the transfer matrix for the Klein-Gordon equation is the same as in the Schr\"odinger case. 
The single notable difference is that now we have 
\begin{equation}
    k_x = \frac{1}{\hbar c}\sqrt{E^2-\hbar^2 c^2 k_y^2-m^2c^4 }
\end{equation}
and
\begin{equation}
    k_x^{\prime} = \frac{1}{\hbar c}\sqrt{(E-V)^2-\hbar^2 c^2 k_y^2-m^2c^4 }.
\end{equation}
In this case, the dependence of the transmission on the incidence angle is given by
\begin{equation}
    k_y = \sin \theta_0 \frac{\sqrt{E^2-m^2c^4}}{\hbar c}.
\end{equation}

\subsection{Results}

\begin{figure}
\centering
\includegraphics[width=0.9\linewidth]{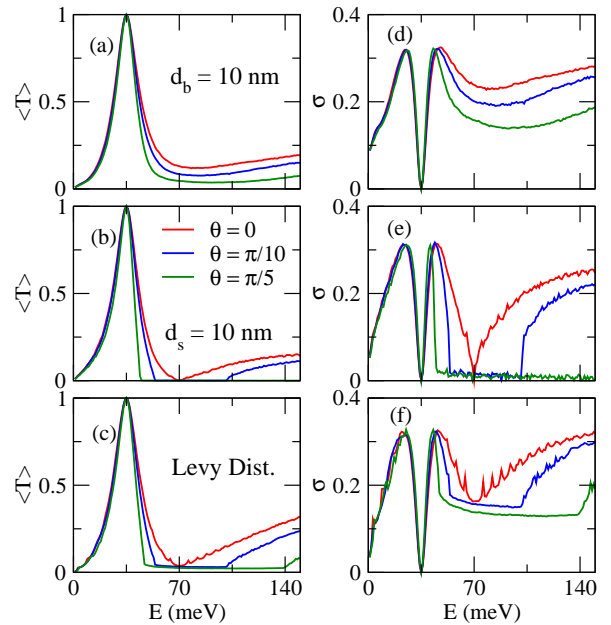}
\caption{Top panels (a,d) correspond to the lattice architecture of Fig. \ref{potencial}(a), middle panels (b,e) to Fig. \ref{potencial} (b) and down panels (c,f) to Fig. \ref{potencial} (c). Transmission average as a function of energy from Klein-Gordon equation for (a)  barrier width $d_b = 10$ nm, (b) barrier separation $d_s = 10$ nm and (c) L\'evy distribution for both. In all calculations we used $\alpha = 0.5$ and potential barrier height $V = 70$ meV for different incidence angles $\theta_0 = 0, \pi/10$ and $\pi/5$. Panels (d-f) are the respective transmission fluctuations as a function of energy.}
\label{Imagem3}
\end{figure}

In order to perform a direct comparison with the results for non-relativistic waves presented in Sec. \ref{sec:schrodinger}, we analyze the same cases considered previously.
Fig. \ref{Imagem3} (a), (b) and (c) show the average transmission $\left\langle T \right\rangle$, while Fig. \ref{Imagem3} (d), (e) and (f) show their respective standard deviation, as a function of energy for the  barrier architectures illustrated in Fig. \ref{potencial}, respectively.
The data in Fig. \ref{Imagem3} has been averaged over $10^4$ realizations with fixed potential barrier height $V=70$ meV, length $L=10$ $\mu$m and L\'evy distribution's exponent $\alpha=0.5$. 
Furthermore, in order to allow a proper  comparison with the Dirac case that will be considered in the next section, we also considered $m=0$ and $c=v_F=10^6$~m/s. 
In the Dirac case, those values are obtained for low-energy electronic excitations in graphene, which is a material that could be used to experimentally verify some of our results \cite{Lima2016,Lima2018,Lima2019}.

It is apparent on the left side panels of Fig. \ref{Imagem3} that the average transmission displays a maximum $\left\langle T \right\rangle=1$ when the energy equals half of the potential barrier height, $E=V/2=35$ meV. 
This behavior is independent of the barrier architecture considered, and it is known as  super-Klein tunneling \cite{Kim2019,PhysRevB.101.165428}. 
After the maximum, the average transmission decreases to a minimum value. 
Depending on the incidence angle $\theta_0$ and the barrier architecture, $\langle T \rangle$ may or may not form a plateau, before increasing again. 
The standard deviation also shows a similar behavior for all barrier architectures. However, it has two outstanding features. First, its maximum value is architecture-independent and equals the value obtained for non-relativistic waves (see Fig. \ref{Imagem1}), which reinforces the universal character of the transmission fluctuations. 
Second, the fluctuation vanishes when $E=V/2=35$ meV, which is also a consequence of the super-Klein tunneling.

The energy at which $\langle T \rangle$ reaches its minimum value shown on the left side panels of Fig. \ref{Imagem3}, the critical energy, can be obtained from the relation
\begin{equation}
    \theta_0=\arcsin \frac{\sqrt{(E_c-V)^2-m^2c^4}}{\sqrt{E_c^2-m^2c^4}}.\label{KG1}
\end{equation}
Taking $m=0$ in the above equation, it simplifies to
\begin{equation}
    \theta_0=\arcsin\left|1-\frac{V}{E_c}\right|, \label{KG}
\end{equation}
and for $\theta_0=0$ we have $E_c=V$. 
However, if $\theta_0>0$, Eq. (\ref{KG})  has two possible values for $E_c$.
The first one is where the transmission plateau in Fig. \ref{Imagem3} (b) and (c) starts, and the second one is where it ends.

\begin{figure}
\centering
\includegraphics[width=0.9\linewidth]{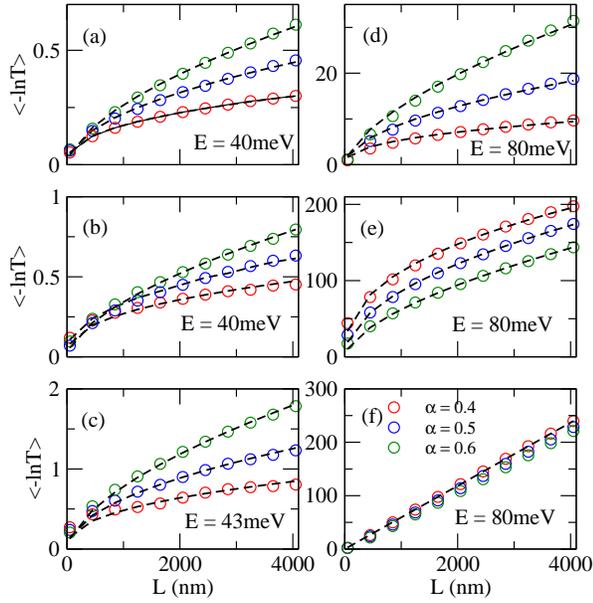}
\caption{The top panels (a,d) correspond to the lattice architecture of Fig. \ref{potencial} (a), middle panels (b,e) to Fig. \ref{potencial} (b) and down panels (c,f) of Fig. \ref{potencial} (c).  $\left\langle -\ln T \right\rangle$ as a function of $L$ for $V = 70$ meV and fix incidence energy when (a-c) $E < E_c$ and (d-f) $E > E_c$. The incidence angle is $\theta_0 = \pi/5$ for top and middle panels and $\theta_0 = \pi/6$ for down panels. In (a-e), the dashed lines are fitted using Eq. (\ref{lnLa}), while in (f) we employ a linear fit.}
\label{Imagem4}
\end{figure}

Let us now analyze the localization regime for a relativistic wave described by the Klein-Gordon equation. 
In Fig. \ref{Imagem4} we plot  $\left\langle - \ln T \right\rangle$ as a function of $L$ for the three barrier architectures considered.
In panels (a), (b) and (c) we have $E<E_c$, whereas $E>E_c$ for the remaining ones.
It is important to mention that, for the cases with $\theta_0 \neq 0$, we consider the lower value of $E_c$  obtained from Eq. (\ref{KG}). 
The numerical data (circles) is well fitted by Eq. (\ref{lnLa}) (dashed lines), which means the system experiences anomalous localization. 
When the energy is increased to $E>E_c$, but without exceeding the second value of the critical energy, which means that the energy is inside the transmission plateau shown in Fig. \ref{Imagem3}, the numerical data of Figs. \ref{Imagem4} (d) and (e) are also fitted by Eq. (\ref{lnLa}), which again is compatible with anomalous localization. 
However, in Fig. \ref{Imagem4} (f) there is a linear increase with $L$, which indicates a  standard localization regime.

Therefore, according to the data in Fig. \ref{Imagem4}, when the potential barriers follow the architectures in Figs. \ref{potencial} (a) and (b) the Klein-Gordon system presents anomalous localization for both $E<E_c$ and $E>E_c$, just like the non-relativistic Schrodinger waves as shown in Fig. \ref{Imagem2}. 
However, when the potential barriers follow the architecture in Fig. \ref{potencial} (c) the Klein-Gordon system presents anomalous localization for $E<E_c$ and standard localization for $E>E_c$, which is the exact opposite of the non-relativistic Schrodinger waves shown in Fig. \ref{Imagem2} (f) and (i). 
This shows that only when the width of the barriers and their separation follow a L\'evy distribution both Klein-Gordon and Schr\"odinger waves present a transition between anomalous and standard localization regimes as the energy increases beyond a critical value.
In the case of non-relativistic Schr\"odinger waves we have a standard-to-anomalous localization transition, whereas in the relativistic Klein-Gordon case we have an anomalous-to-standard localization transition. 

\section{Dirac equation}
\label{sec:Dirac}

\subsection{Model}

The Dirac Hamiltonian in two dimensions for fermions subject to one-dimensional potential barriers as shown in Fig. \ref{potencial} can be written as
\begin{eqnarray}
H=-i\hbar c \left(\sigma_x \partial_x +\sigma_y \partial_y\right) + mc^2\sigma_z
+V(x),
\end{eqnarray}
where $\sigma_i$ are the Pauli matrices. 
The Dirac equation is written as
\begin{equation}
H\psi(x,y)=E\psi(x,y),
\end{equation}
where $\psi(x,y)$ is a two-component spinor, which we can write in the form $\psi(x,y)=(\psi_A(x,y), \psi_B(x,y))^T$. Writing $\psi(x,y)=e^{ik_yy}\psi(x)$, the Dirac equation becomes
\begin{eqnarray}
[-i\hbar c(\sigma_x\partial_x + ik_y \sigma_y) +mc^2\sigma_z]\psi(x)  =(E-V)\psi(x) \; ,
\end{eqnarray}
which can be recast as
\begin{equation}
i\frac{d\psi(x)}{dx}=P(x)\psi(x) \; ,
\label{dirac}
\end{equation}
where
\begin{equation}
P(x)=\left(
\begin{array}{cc}
ik_y & \frac{-(E-V(x))-mc^2}{\hbar c} \\
\frac{-(E-V(x))+mc^2}{\hbar c} & -ik_y
\end{array} \right) \; .
\end{equation}
Inside the $j$th region $V(x)$ is constant and one can write 
\begin{equation}
\frac{d^2\psi_{A,B}}{dx^2} + (k^2_j - k^2_y) \psi_{A,B} = 0  ,
\end{equation}
where $k_j = [(E-V_j)^2-m^2c^4]^{1/2}/(\hbar c)$ is the wave vector.

Following the calculations in [\onlinecite{PhysRevB.81.205444}], we can obtain the transfer matrix connecting the wave function at $x$ and $x+\Delta x$ in the $j$th region as
\begin{equation}
M_j(\Delta x, E, k_y)=\left(
\begin{array}{cc}
\frac{\cos(q_j\Delta x - \theta_j)}{\cos\theta_j} & i\frac{\sin(q_j\Delta x)}{p_j \cos\theta_j} \\
i\frac{p_j \sin(q_j\Delta x)}{\cos\theta_j} & \frac{\cos(q_j\Delta x + \theta_j)}{\cos\theta_j}
\end{array} \right) \; ,
\end{equation}
where $p_j=l_j/k_j$ with $l_j=[(E-V_j)-m c^2]/(\hbar c)$. 
The quantity $q_j$ is the $x$ component of the wave vector, given by $q_j = \sqrt{k_j^2-k_y^2}$ for $k_j^2>k_y^2$, otherwise $q_j = i\sqrt{k_y^2-k_j^2}$.  $\theta_j$ is the angle between the $x$ component of the wave vector, $q_j$, and the wave vector $k_j$ and it is given by $\theta_j = \arcsin (k_y/k_j)$. 
So, the transfer matrix that connects the incident and outgoing wave functions for a superlattice with N regions is given by
\begin{equation}
X = \left(
\begin{array}{cc}
x_{11} &  x_{12}\\
x_{21} & x_{22}
\end{array} \right) = \prod_{j=1}^{N} M_j(w_j, E, k_y).
\end{equation}
where $w_j$ is the width of the $j$th region.

Finally, the transmission coefficient is given by
\begin{equation}
t(E, k_y) = \frac{2\cos \theta_0}{(x_{22}e^{-i\theta_0}+x_{11}e^{-i\theta_e})-x_{12}e^{i(\theta_e-\theta_0)}-x_{21}},
\label{tc}
\end{equation}
where $\theta_0$ and $\theta_e$ are the incidence and exit angles, respectively. 

\begin{figure}
\centering
\includegraphics[width=0.9\linewidth]{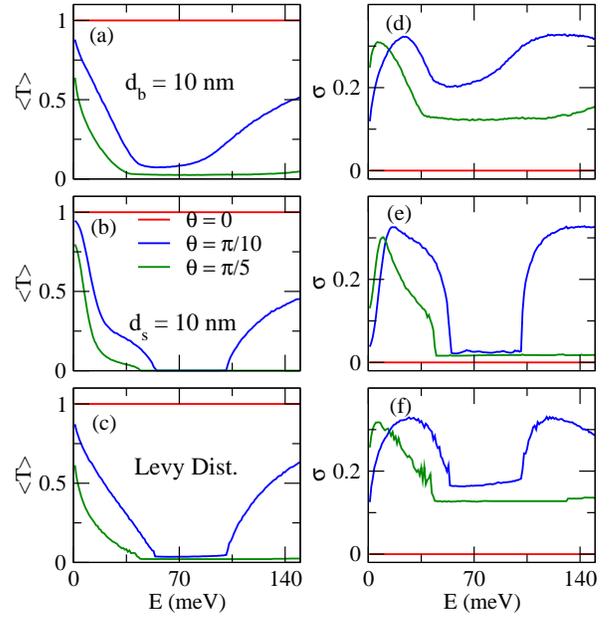}
\caption{Top panels (a,d) correspond to the lattice architecture of Fig. \ref{potencial}(a), middle panels (b,e) to Fig. \ref{potencial} (b) and down panels (c,f) to Fig. \ref{potencial} (c). Transmission average as a function of energy from Dirac equation for (a)  barrier width $d_b = 10$ nm, (b) barrier separation $d_s = 10$ nm and (c) L\'evy distribution for both. In all calculations we used $\alpha = 0.5$ and potential barrier height $V = 70$ meV for different incidence angles $\theta_0 = 0, \pi/10$ and $\pi/5$. Panels (d-f) are the respective transmission fluctuation as a function of energy.}
\label{Imagem5}
\end{figure}

\subsection{Results} 
Following the same steps employed in the Schr\"odinger and Klein-Gordon cases, we analyze now the localization regime of relativistic waves described by the Dirac equation.
The left side panels of Fig. \ref{Imagem5} show the average transmission $\left\langle T \right\rangle$, while the right side panels show the corresponding standard deviation, as a function of energy for the three barrier architectures represented in Fig.\ref{potencial}.
Once again the data has been averaged over $10^4$ independent realizations, with fixed potential $V=70$ meV, $L=10$ $\mu$m and L\'evy distribution exponent $\alpha=0.5$. 
We also choose the parameters $m=0$ and $c=v_F=10^6$~m/s as in graphene, in order facilitate future experimental verification.

\begin{figure}
\centering
\includegraphics[width=0.9\linewidth]{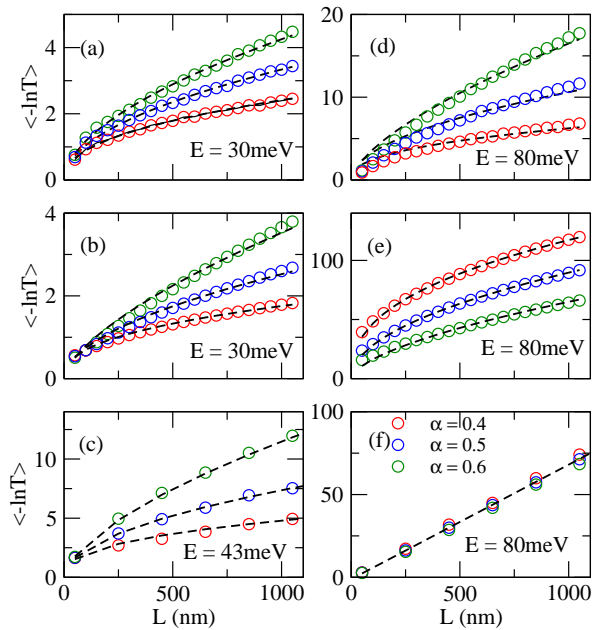}
\caption{The top panels (a,d) correspond to the lattice architecture of Fig. \ref{potencial} (a), middle panels (b,e) to Fig. \ref{potencial} (b) and down panels (c,f) of Fig. \ref{potencial} (c).  $\left\langle -\ln T \right\rangle$ as a function of $L$ for $V = 70$ meV and fix incidence energy when (a-c) $E < E_c$ and (d-f) $E > E_c$. The incidence angle is $\theta_0 = \pi/5$ for top and middle panels and $\theta_0 = \pi/6$ for bottom panels. In (a-e), the dashed lines are fitted using Eq. (\ref{lnLa}), while in (f) we employ a linear fit.}
\label{Imagem6}
\end{figure}

One remarkable feature in Fig. \ref{Imagem5} is the presence of Klein tunneling at normal incidence $\theta_0=0$, i.e. perfect transmission with zero standard deviation, independent of the barrier architecture and energy.
Meanwhile, for $\theta_0>0$ the average transmission decreases and reaches a minimum, which becomes a plateau in some cases, after which the transmission increases again. 
The standard deviation presents one or more maxima, which has the same value for the three lattice architectures and also for the Schr\"odinger and Klein-Gordon cases, which again indicates a certain  universality for transmission fluctuations.
The critical energy for the Dirac case is given by the same expression obtained in the Klein-Gordon case, i.e.  Eqs. (\ref{KG1}) and (\ref{KG}). 

As in the previous cases, let us now study the localization behavior of relativistic waves described by the  Dirac equation for the three barrier architectures illustrated in Fig. \ref{potencial}. 
In Fig. \ref{Imagem6} panels (a-c) present $\langle - \ln T \rangle$ as a function of $L$ when $E<E_c$ for the three barrier architectures. 
Again, we take $E_c$ to be the lower solution in Eqs. (\ref{KG1}) and (\ref{KG}). 
The numerical data (circles) is well fitted by Eq. (\ref{lnLa}) (dashed lines), which again is compatible with anomalous localization. 
When the energy is increased beyond the critical value, $E>E_c$, the data in panels (d) and (e) is also fitted by Eq. (\ref{lnLa}), and thus compatible with anomalous localization. 
However, in panel (f) $\langle -\ln T \rangle$ shows a linear dependence with $L$, which is only compatible with standard localization. 
Note that the localization behavior of waves described by the Dirac equation equals the one we found for the Klein-Gordon case in Sec. \ref{sec:KG}, and it's the opposite of the Schr\"odinger case in Sec. \ref{sec:schrodinger}.

Finally, from Figs. \ref{Imagem2}, \ref{Imagem4} and \ref{Imagem6} we conclude that the potential barrier architectures in Fig. \ref{potencial} (a) and (b) induce anomalous localization for both $E<E_c$ and $E>E_c$, no matter if the description is non-relativistic or relativistic. 
However, the barrier architecture in Fig. \ref{potencial} (c), induces standard localization when $E<E_c$ and anomalous localization when $E>E_c$ for non-relativistic waves, and the opposite for relativistic waves i.e. anomalous localization for $E<E_c$ and standard localization for $E>E_c$.

\section{Discussion}

\begin{figure}[htb]
\centering
\includegraphics[width=0.9\linewidth]{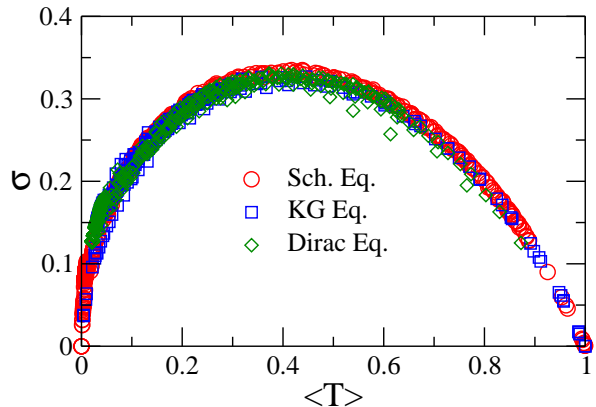}
\caption{Transmission standard deviation $\sigma$ as a function of average transmission for Schr\"odinger, Klein-Gordon and Dirac equations with $\alpha=0.4, 0.5$ and 0.6. The figure shows a universal behavior independent of lattice architecture, the L\'evy exponent $\alpha$ and model equation.}
\label{Imagem7}
\end{figure}

The results presented so far show a transition from standard to anomalous localization, or vice versa, when a L\'evy-type disorder is present for both the width of the potential barriers and the separation between them.
Nonetheless, if we plot the standard deviation of the transmission versus the average transmission, we find a universal behavior regardless of the specific equation describing the propagation, the type of disorder in the superlattice, and the L\'evy exponent $\alpha$, as shown in Fig. \ref{Imagem7}.
When the transmission is close to its lowest or highest values, the  fluctuations are relatively small.
Meanwhile, when $\langle T \rangle$ assumes intermediate values its fluctuations reach a maximum.
{The observed results do not depend on the potential barrier height. Indeed, we fixed the potential value at V = 70 meV only to facilitate comparison between the relativistic and non-relativistic equations.}

The result in Fig. \ref{Imagem7} are an indication that both quantities depend on the magnitude of the mean free path and the system length, but do not depend explicitly on the distribution of potential barriers.
This type of universal behavior has been observed since the early days of electron localization studies \cite{Lee1985,Giordano1988}. 
Our results extend this universality beyond  disorder architecture to also include the description of the waves by relativistic or non-relativistic equations.
Analogous behavior has been observed for phonon transport in superlattices, both analytically and numerically \cite{Nishigushi1993}.

\begin{figure*}[htb]
\centering
\includegraphics[width=0.7\linewidth]{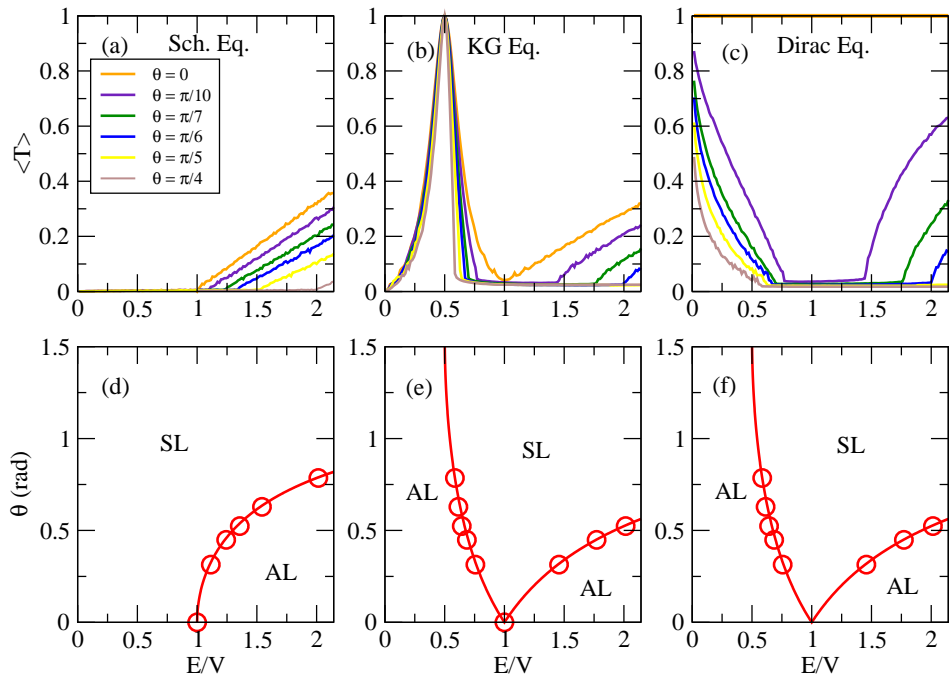}
\caption{Transmission average as a function of {ratio between incidence energy and potential $E/V$} for superlattices with L\'evy-type disorder with $\alpha=0.5$ for both barriers and their separation, as shown in Fig. \ref{potencial} (c). 
(a) Schr\"odinger (b)  Klein-Gordon (c) Dirac equations and different incidence angles. 
Bottom panels (d-f) show the localization diagram in terms of incidence angle versus {$E/V$}, dividing anomalous (AL) and standard localization (SL) regimes, for Schr\"odinger, Klein-Gordon and Dirac equations respectively. The localization diagrams are valid for all L\'evy exponents $\alpha$ between 0 and 1.}

\label{fig09}
\end{figure*}

Finally, we now present the localization diagram in the presence of L\'evy-type disorder with $\alpha=0.5$ for barriers and their separation, the architecture shown in Fig. \ref{potencial} (c).
In the top panels of Fig. \ref{fig09} we show the transmission as a function of {ratio between energy and potential $E/V$} for the three equations considered here and six incidence angles, while the bottom panels show the energy dependence of the critical angle for each one of the equations.
The localization diagrams shown in the bottom panels of Fig. \ref{fig09} are valid for all L\'evy exponents $\alpha$ between 0 and 1.
In the non-relativistic case there are only two regions in the  diagram: standard and anomalous localization. 
Meanwhile, in the relativistic cases, there are three regions: one for standard localization and two for anomalous localization.
Therefore, for a fixed incidence angle in the non-relativistic case there is only one standard-to-anomalous localization transition as the energy increases, whereas in the relativistic cases there can be two transitions, accounting for an anomalous-to-standard-to-anomalous localization transition.
The presence of the anomalous localization regime in the low-energy region of the relativistic cases is most likely related to Klein tunneling and super-Klein tunneling shown in Secs. \ref{sec:KG} and \ref{sec:Dirac}, which are not present in the non-relativistic case.

Finally, for the standard localization cases, we estimate from Figs. \ref{Imagem2}(f), \ref{Imagem4}(f) and \ref{Imagem6}(f) that the localization lengths are $\lambda=2.7$ nm for the Sch\"odinger case, $\lambda=17.5$ nm for the Klein-Gordon case, and $\lambda=14.5$ nm for the Dirac case. 
The localization lengths for the relativistic cases are one order of magnitude larger than the non-relativistic one. 
A shorter localization length for non-relativistic systems is consistent with a previous study which investigated the influence of relativistic effects on localization in disordered systems, considering the Schr\"odinger and Dirac equations \cite{Basu1994}.
On the other hand, our results show that the relativistic cases have similar localization lengths.

\section{Conclusion}

In conclusion, we employed the transfer matrix method to analyze non-relativistic and relativistic wave transmission submitted to potential barriers whose width, separation or both follow L\'evy distributions.
We found that for the first two cases, where one of the parameters was fixed, non-relativistic and relativistic waves present anomalous localization.
However, for the latter case, in which both parameters follow a L\'evy distribution, we found that non-relativistic waves present a standard-to-anomalous localization transition, while relativistic waves present an anomalous-to-standard-to-anomalous transition.
We obtained the localization diagram delimiting anomalous and standard localization regimes for each case, in the form of an incidence angle versus incidence energy diagram. 
The presence of the anomalous localization regime in the low-energy region of the relativistic cases can be attributed to Klein tunneling and super-Klein tunneling, which are absent in the non-relativistic case. 
Finally, we verified that transmission fluctuations (characterized by its standard deviation) are universal, independent of barrier architecture, wave equation type, incidence energy and angle, further extending earlier studies on localization of electrons and phonons.

\acknowledgments

ALRB acknowledges financial support from Conselho Nacional de Desenvolvimento Científico e Tecnológico - CNPq (Grant 309457/2021-1).
JRFL acknowledges CNPq for financial support (Grant 316179/2021-3).  
LFCP acknowledges financial support from CNPq (Grants 309961/2017, 436859/2018 and 313462/2020).

\bibliography{library}

\end{document}